\documentclass[amssymb,superscriptaddress,twocolumn,prd]{revtex4}
\usepackage{graphicx}
\def \be{\begin{equation}}
\def \ee{\end{equation}}
\def \bea{\begin{eqnarray}}
\def \eea{\end{eqnarray}}

\newcommand{\beq}[1]{\begin{eqnarray}\label{#1}}
\newcommand{\eeq}{\end{eqnarray}}

\begin{document}
 \pagestyle{plain}

 \title{An Ignored Assumption of $\Lambda$CDM Cosmology
 \\ and An Old Question:
 \\ Do We Live On The ``Center'' of The Universe?}

\author{Ding-fang Zeng and Yi-hong Gao}
\email{dfzeng@itp.ac.cn} \email{gaoyh@itp.ac.cn}
\affiliation{Institute of Theoretical Physics, Chinese Academy of
Science.}
 \begin{abstract}
 We point out that $\Lambda$CDM cosmology has an ignored
 assumption. That is, the $\Lambda$ component of the universe
 moves synchronously with ordinary matters on Hubble scales. If cosmological
 constant is vacuum energy, this assumption may be very difficult
 to be understood.

 We then propose
 a new mechanism which can explain the accelerating recession of
 super-novaes. That is, considering the pressures originating from
 the random moving (including Hubble recession) of galaxy clusters
 and galaxies. We provide an new analytical solution of Einstein
 equation which may describe a universe whose pressures
 originating from the random moving of galaxy clusters and
 galaxies are considered.
 \end{abstract}


 \maketitle

 \section{The Ignored Assumption of $\Lambda$CDM Cosmology}
 $\Lambda$CDM Cosmology has an ignored assumption. That is,
 the $\Lambda$ component of the universe moves synchronously with
 ordinary matters on Hubble scales.
 Usually, this assumption is made when
 we write down the Einstein equation
 $G_{\mu\nu}=-8\pi GT_{\mu\nu}-\Lambda g_{\mu\nu}$
 to describe the evolution of the
 universe. Since otherwise, in the co-moving reference frame
 which is set up on the matter component,
 we will have an $\Lambda$ current
 flowing inside the Hubble horizon of our
 universe as it expands.
 Under this case, the energy momentum cannot be
 diagonalized, so we can prove that for the metric ansaltz,
 $ds^2=-dt^2+U(t,r)dr^2+V(t,r)(d\theta^2+\textrm{sin}^2\theta
 d\phi^2)$, the functions $U(t,r)$ and
 $V(t,r)$ cannot be globally factorized as $a^2(t)\cdot f(r)$ and $a^2(t)\cdot r^2$
 respectively, so we have no a globally defined
 scale factor at all, and no Friedmann equation either. For
 details of the argument in this paragraph,
 please refer to \cite{SWeinberg},
 sections. 11.8-9, \cite{SphereI} and \cite{SphereII}.

 If $\Lambda$ component is vacuum energy, then we have no reason
 to say that it moves synchronously with ordinary matters, even
 on the Hubble scales.
 On the contrary, the conception that it acts as an absolute background
 with negative pressures may be more favored by our intuitions.
 However, if $\Lambda$ component is vacuum energy and does not move
 synchronously with ordinary matters, then
 a serious problem must be notified. That is,
 our current explanation
 of the accelerating expansion of the universe should be
 re-considered
 because it is based on the assumption that $\Lambda$ component
 moves synchronously with ordinary matters.
 As we state in the previous paragraph, when
 the $\Lambda$ current is included in the energy momentum tensor,
 the scale factor of the universe cannot be defined globally,
 i.e.,we can define scale
 factors $a_1$ and $a_2$ through relations $U(t,r_1)=a_1(t)\cdot f(r_1)$ and $U(t,r_2)=a_2(t)\cdot
 f(r_2)$ respectively, but we cannot assure that $a_1(t)=a_2(t)$.
 In this case, the explanation of the accelerating expansion of the universe
 must be very different from that in the usual $\Lambda$CDM
 cosmology where the scale factor of the universe is defined
 globally.

 If the vacuum energy of quantum field must be considered in
 cosmology and it is required to move synchronously with ordinary
 matters, we guess but are not sure some kinds of
 regularization is necessary. As
 we know, in the studying of Casimir effects between two parallel
 conducting plates, the vacuum energy is regularized, with
 negative pressures and is of
 course strictly co-moving with our conducting plates. But we do
 not know when the regularization is applied on our universe, it
 is possible or not a vacuum energy density of
 order $G^{-1}H_0^2$ which is required by astro-physical
 observations can be obtained. Whatever order the regularized vacuum
 energy density is of, we think when the regularization is
 performed, the cosmological constant problem
 should not be so serious as we usually think.

 From the following Gendanken experiment, we can also see that the
 vacuum energy which does not co-move with ordinary matters cannot
 contribute
 to the accelerating expansion of the universe. First imagine that our
 universe only consists of two test particles and an omnipresent
 vacuum energy. For each of the test particles, the repulsion
 force comes from the different direction of the background vacuum
 energy cancels each other as long as the background space time is
 infinitely large. So our two test particles can only move towards
 each other at the inter-gravitations. Then imagine that our universe
 consists only of a galaxy clusters in which our Milky Way galaxy
 lies in and an omnipresent vacuum energy. In this case, the
 repulsion exerted on the galaxies inside the clusters
 comes from the different direction of the background vacuum energy
 also cancels each other, and the galaxies only move towards the
 center of the clusters at the galaxy-galaxy-between gravitations.
 Depends on the relative value of the initial kinetic energy and
 the inter-gravitating potential of the galaxies, our galaxy
 cluster as a whole can be at accelerating contraction phase
 accelerating expansion phase. But its contraction or expansion phase
 has nothing to do with the omnipresent vacuum energy. It is
 the pressures originated from the random moving of the galaxies that
 determines the phase our galaxy cluster lies on.
 Finally let us imagine that, our universe
 consists of many many galaxy clusters and galaxies and the
 omnipresent vacuum energy, the only difference of this case from
 that of the one galaxy-cluster universe is that, when we are
 considering the cluster-cluster, cluster-galaxy and galaxy-galaxy
 interactions, we only need to count contributions from those clusters and
 galaxies lie in the particle horizon of the universe.

 Further analyzing our Gedanken experiment of one galaxy cluster
 universe, we doubt we may forget something of very importance when we
 extrapolate the concepts of pressures obtained from the
 ideal gas in laboratories to the
 expanding universe. To make our doubt more clearly expressed, let
 us consider three kinds of gases, (i) a bottle of closed gas in
 our laboratories, (ii) a galaxy cluster as a gas whose basic
 molecules are galaxies such as our Milky Way and (iii) the total
 observable universe as gas whose basic molecules are galaxy
 clusters and galaxies. In the first case, the pressure of the gases
 originates from the random moving of the basic molecules and
 the system is kept in bounding state because the bottle has
 walls which cannot be penetrated by the basic molecules. In the second case, the
 pressure also originates from the random moving of the basic
 molecules - the galaxies, the system is kept in balance at the
 self-gravitation and the pressures. In the third case, we are
 usually told that if dark energy were not introduced, the
 cosmic fluid is a zero-pressure gas. Is it reasonable to neglect
 the pressures originating the random moving of galaxy clusters and galaxies?

 We must note that it is just the same pressures
 that kept the galaxy clusters in a stable state in the second example.
 In the second example, if initially, the composite galaxies were
 given too much kinetic energy, then the total
 system may not be able to be kept in balance, some of its
 composite molecules may escape, just as man-made satellites
 escape from our planet. Some people may say that what we state
 here is just the same thing occurs in a totally matter dominated
 open universe. We will say to these peoples that, partly this is
 the case, but it cannot be totally think so. What we would like
 to emphasize here is that, even in a totally matter dominated
 universe, the pressures originating from the random moving of the
 composite molecules of it cannot be neglected as we are usually
 taught. What's more, in a homogeneous and isotropic universe,
 if the Hubble recession of galaxies are also count as random
 movings and also contribute to pressures, we will expect that the
 energy momentum tensor describing our cosmic fluid should have
 such a property that, its pressure part (the abstract value)
 should increase as a
 function of the co-moving distance. In the next section, we will
 provide a strict solution of Einstein equation, whose energy
 momentum tensor indeed has this property, please see
 eqs(\ref{metricAnsaltz}) and (\ref{EnergyMomentumTensor}).

 \section{A New Mechanism Explaining the Accelerating Recession of Super-Novaes}

 At this moment, we are not
 sure that the existence of particle horizon guarantees the
 regularization of vacuum energy and the regularized vacuum
 energy will affect the expansion of our observable universe.
 So, rather than integrating the vacuum energy of
 quantum field and regularizing it to get the cosmological
 constant, we prefer to propose another mechanism which can
 explain the accelerating expansion of our universe and which is
 of no necessary to introduce the $\Lambda$ term in the Einstein
 equation. That is, the pressures originating from the
 random moving of galaxy clusters and galaxies cannot be
 neglected.

 We find that the following metric ansaltz solve Einstein equation
 $G_{\mu\nu}=-8\pi GT_{\mu\nu}$,
 \beq{}
 ds^2=-dt^2+e^{rt/A}(dr^2+r^2d\theta^2+r^2\textrm{sin}^2\theta
 d\phi^2)
 \label{metricAnsaltz}
 \eeq
 with the energy momentum tensor given by
 \beq{}
 &&\hspace{-3mm}8\pi GT_{\mu\nu}=\nonumber\\
 &&\hspace{-3mm}\left[
 \begin{array}{cccc}
 \frac{3r^2-e^{-rt/A}t^2}{4A^2}-\frac{2e^{-rt/A}t}{Ar}&-A^{-1}&0&0\\
 -A^{-1}&\frac{t^2-3e^{rt/A}r^2}{4A^2}+\frac{t}{Ar}&0&0\\
 0&0&T_{22}&0\\
 0&0&0&T_{33}
 \end{array}
 \right]\nonumber\\
 &&\hspace{10mm}T_{22}=\frac{rt}{2A}-\frac{3e^{rt/A}r^4}{4A^2},
 T_{33}=T_{22}\textrm{sin}^2\theta
 \label{EnergyMomentumTensor}
 \eeq
 Where $A$ is a constant with dimension of $[length]^2$.
 Comparing eq(\ref{metricAnsaltz}) with the usual
 Friedman-Robertson-Walker metric $ds^2=-dt^2+a^2(t)(dr^2+r^2d\theta^2+r^2\textrm{sin}^2\theta
 d\phi^2)$, we can see that the most remarkable feature of
 the metric (\ref{metricAnsaltz}) is that, the scale factor is $r$
 dependent and at a given position $r$, the scale factor increase
 exponentially as time passes by. So such a metric can describe an
 accelerating universe. In fact it is just using being able to
 describe an accelerating universe as a criteria that we construct
 the metric ansaltz eq(\ref{metricAnsaltz}), and then using
 Einstein equation we get the energy momentum tensor
 eq(\ref{EnergyMomentumTensor}).

 So at this moment, we are not very familiar with the physical
 meaning of each term in the energy momentum tensor
 eq(\ref{EnergyMomentumTensor}). But we
 know it describes some kinds of fluid whose pressures
 cannot be neglected, and (the abstract value) is an
 increasing function of the co-moving
 distance increases. This point coincides with our expectation
 expressed at the end of previous section. So we guess it may have
 relevance with our realistic cosmological fluid, when the pressures
 originating from the random moving (including the Hubble recession)
 of galaxy clusters and galaxies must be considered.
 Its non-diagonal terms only appear on the
 $t-r$ and $r-t$ position, this may indicate the fact that the
 pressures in cosmology has differences from that of the gas in
 our laboratories. In the studying of the gas pressures in
 laboratories, we need not to consider the time for a gas molecule
 to run from the container's walls to another molecule after collision
 with the container walls.
 But in cosmologies, to consider the effects of pressures
 originating from the random moving or Hubble recessions of galaxy
 clusters and galaxies, the time for a basic composite galaxy to
 run from the Horizon edge to us must be considered. In this case
 we expect that the energy momentum tensor must have non-zero
 $t-r$ and $r-t$ components.

 Of course, since the metric (\ref{metricAnsaltz}) deviates from
 the standard Freidmann-Robertson-Walk metric so much that, we are
 very not sure that it describes our real universe indeed. But we
 are sure that, if the pressures originating from the random
 moving (including the Hubble recession) of galaxy clusters and
 galaxies in our universe must be considered, then the metric of
 our universe must have the same features as
 eq(\ref{metricAnsaltz}), its scale
 factor is both $t$ and $r$ dependent, and this dependence cannot be
 factorized as $a^2(t)f(r)$ as we are taught in the standard text-book
 such as \cite{SWeinberg}. We are
 also sure that, in this case the energy-momentum tensor is
 non-diagonal, it has non-zero $t-r$ and $r-t$ components and its
 energy density and pressures are both time and position
 dependent. For example we can check that the following metric
 ansaltz also solve Einstein equation:
 \beq{}
 ds^2=-dt^2+e^{t/r}(dr^2+r^2d\theta^2+r^2\textrm{sin}^2\theta
 d\phi^2)\label{metricAnsaltz2}
 \eeq
 but with the energy momentum tensor written as
 \beq{}
 &&\hspace{-3mm}8\pi GT_{\mu\nu}=\left[
 \begin{array}{cccc}
 \frac{3-r^{-2}t^2}{4}&r^{-2}&0&0\\
 r^{-2}&-\frac{3e^{t/r}+4r^{-1}t-r^{-2}t^2}{4r^2}&0&0\\
 0&0&T_{22}&0\\
 0&0&0&T_{33}
 \end{array}
 \right]\nonumber\\
 &&\hspace{10mm}T_{22}=-\frac{3e^{t/r}-2r^{-1}t}{4},
 T_{33}=T_{22}\textrm{sin}^2\theta
 \label{EnergyMomentumTensor2}
 \eeq
 We are not sure if this metric describe an accelerating expanding
 universe or not. Because, naively looking, the recession velocity
 of an object in such a space-time reads:
 $\frac{\dot{a}}{a}\propto\frac{1}{r}$, the farer is an object
 lies away from the origin, the smaller its recession velocity
 will be. But, in this space time, since the scale factor of universe
 is $r$ dependent, the definition Hubble recession's velocity may
 not be the same as that in the usual Friedmann-Robertson-Walker
 space-time.

 We must emphasize again that, the metrics eq(\ref{metricAnsaltz})
 and (\ref{metricAnsaltz2}) is obtained by an inverse method. We
 are not sure it can describe a realistic universe. We provide it
 here only to illustrate that if the pressures originating from
 the random moving of galaxy clusters and galaxies in the universe
 cannot be neglected, the metric of the space time and the energy momentum
 tensor should have the same features as those of
 eqs(\ref{metricAnsaltz}), (\ref{EnergyMomentumTensor}) and
 eqs(\ref{metricAnsaltz2}), (\ref{EnergyMomentumTensor2}).

 Originally, we think that to get a solution of Einstein equation in which
 the scale factor of metric is both time and position dependent
 and if the dependence cannot be factorized, so that we can have a
 position dependent Hubble recession, our observable universe must
 be lying on the center of some very big super-super-clusters, in
 this super-super-cluster, the matter distribution has spherical
 symmetry, so in the general solution of Einstein equation
 $ds^2=-dt^2+U(t,r)dr^2+V(t,r)d\Omega_2^2$, the function $U(t,r)$ cannot be
 factorized as $a^2(t)f(r)$ as we are usually taught in the
 text-book such as \cite{SWeinberg}, we test the ansaltz
 eq(\ref{metricAnsaltz2}) and get the energy momentum tensor
 eq(\ref{EnergyMomentumTensor2}). Although we find that, the
 energy density and pressures are decreasing function of co-moving
 distances, but the metric may not describe an accelerating
 universe as indicated by the super-novae observations. Then we
 test metric ansaltz eq(\ref{metricAnsaltz}) and get the energy
 momentum tensor eq(\ref{EnergyMomentumTensor}), in this case we
 find the metric describe an accelerating expansion universe, but
 the energy-density and pressures is an increasing function of
 co-moving distances. In this case, we propose that in the usually
 Friedmann-Robert-Walker universe, the pressures originating from
 the random moving of galaxy clusters and galaxies may be
 neglected un-appropriately. If our original imagination were
 correct, we may really live on the center of the ``universe''. so
 we name our this paper as
 \newline``An Ignored Assumption of $\Lambda$CDM Cosmology
 and An Old Question: Do We Live On The Center of The
 ``Universe''?''
 \newline
 Although we have realized that, as long as we consider the pressures
 originating from the random moving (including the Hubble recession)
 of galaxy clusters and galaxies,
 obtaining an non-factorized scale factor of the universe is
 possible even discarding the assumption that we are living on the center of the
 ``universe'', we still would like to name this paper
 as
 \newline
 ``An Ignored Assumption of $\Lambda$CDM Cosmology
 and An Old Question: Do We Live On The ``Center'' of The
 Universe?''
 \newline
 but now with the position of quotation mark changed!
 Because from some aspects, we are living on a
 minimum pressure point in the universe, it may be thought as the
 ``center'' of the universe!

 \section{Conclusions}
 In the first
 part of this paper, we point out that $\Lambda$CDM cosmology has
 an ignored assumption. That is, the $\Lambda$ component of the
 universe moves synchronously on Hubble scales. We think this is
 very difficult to understand if it is vacuum energy of quantum
 field. But, considering that in the studying of the Casimir
 effects of two parallel conducting plates, after regularization,
 the vacuum energy obtained is not only co-moving with the
 conducting plates, but also with negative pressures, we think
 that to get the correct cosmological constant by integrating the
 vacuum energy, some kind of regularization is necessary.

 In the second part of the paper, we propose a new mechanism which
 can explain the accelerating recession of super-novaes but with
 no necessary to introduce $\Lambda$ term in the Einstein
 equation. In this mechanism, the pressures of the cosmological
 fluid originating from the random moving of galaxy clusters and
 galaxies is considered.
 We provide an analytical solution of Einstein equation which may
 describe some kinds of cosmological fluid whose pressures is
 considered. We are not very sure that our metric describes our
 realistic universe indeed. But from the metric we can see that,
 if the pressures originating from the random moving of
 galaxy-clusters and galaxies must be considered, the scale factor
 of the universe metric should depend on the time and positions at
 the same time, and the dependence on the two variables cannot be
 factorized. At the same time, the energy-momentum tensor of the
 universe should not be diagonal, it should have non-zero
 component on the $t-r$ and $r-t$ position to indicate the fact
 that, for a test particle to move from the edge of the horizon to
 us, time is needed.

\end{document}